\documentclass[prl,twocolumn,amsmath,amssymb,graphicx]{revtex4-2}
\usepackage{bm,url,natbib,textcase,color,hyperref}
\usepackage[utf8]{inputenc}
\usepackage{amsmath,amsthm,amssymb,mathtools,tikz,booktabs,framed,eucal}

\newcommand{\mc}{\mathcal}  \newcommand{\mb}{\mathbb} \newcommand{\on}{\operatorname}  

\newcommand{\eprinta}[1]{{\href{http://arxiv.org/abs/#1}{[\texttt{#1}]}}}

\newcommand{\eprintN}[1]{{\href{http://arxiv.org/abs/hep-th/#1}{[\texttt{#1 [hep-th]}]}}}

\newcommand{\eprintMath}[1]{{\href{https://arxiv.org/abs/math/#1}{[\texttt{#1}]}}}
\newcommand{\doia}[2]{{\hypersetup{urlcolor=blue}\href{http://dx.doi.org/#2}{#1}\hypersetup{urlcolor=blue}}}

\begin{document}

{\phantom{.}\vspace{-2.5cm}\\\flushright Imperial-TP-KM-2022-03\\ }
\bigskip

\title{Actions for type II supergravities and democracy}

\author{Karapet Mkrtchyan} 
\email{k.mkrtchyan@imperial.ac.uk}
\author{Fridrich Valach} 
\email{f.valach@imperial.ac.uk}

\affiliation{Theoretical Physics Group, Blackett Laboratory, Imperial College London, SW7 2AZ, UK}

\begin{abstract}

We present a universal democratic Lagrangian for the bosonic sector of ten-dimensional type II supergravities, treating ``electric'' and ``magnetic'' potentials of all RR fields on equal footing. For type IIB, this includes the five-form whose self-duality equation is derived from the Lagrangian. We also present an alternative form of the action for type IIB, with manifest $SL(2,\mb R)$ symmetry.

\end{abstract}

\maketitle

\section{Introduction}


A large part of our knowledge on string theory comes from its low-energy limit --- 10d supergravities. The rich symmetry structures of the latter theories are governed by generalized geometry \cite{Hitchin:2003cxu,Gualtieri:2003dx,Grana:2005sn,Grana:2008yw,Coimbra:2011nw} (see also \cite{Siegel:1993xq,Siegel:1993th}; for a review, see, e.g., \cite{Plauschinn:2018wbo}) operating not only with metric tensor but also form fields. It is therefore very useful to have descriptions of supergravities that manifest large number of underlying symmetries of string theory. We start investigation in this direction by developing a new Lagrangian formulation for type II supergravities adapted to generalised geometry. That is, they treat the electric and magnetic degrees of freedom in equal footing for the RR sector. This formulation we put forward in this work comes with a large number of new symmetries that will be explored in more detail elsewhere. 

The action formulation for type IIB supergravity is not straightforward due to the presence of the self-dual field \cite{Witten:1996hc,Witten:1999vg,Sen:2015uaa} and many treatments \cite{Schwarz:1983qr,Howe:1983sra,Castellani:1993ye,Bergshoeff:1995as,Townsend:1996xj,Coimbra:2011nw,Ciceri:2014wya,Bossard:2021jix,Bossard:2021ebg} use the so-called pseudoaction, where one imposes the self-duality equation on top of the Euler-Lagrange equations. The (twisted) self-duality condition is known to be hard to incorporate in the Lagrangian theory, even a free one. While some formulations of self-dual fields depart from manifest Lorentz covariance \cite{Marcus:1982yu,Floreanini:1987as,Henneaux:1988gg,Tseytlin:1990nb,Schwarz:1993vs}, several covariant approaches were developed \cite{Siegel:1983es,McClain:1990sx,Bengtsson:1996fm,Pasti:1996vs,Devecchi:1996cp,Rocek:1997hi,Sen:2019qit,Mkrtchyan:2019opf} and Lagrangians for type IIB supergravity were introduced using these approaches \cite{Kavalov:1986ki,Berkovits:1996nq,DallAgata:1997gnw,DallAgata:1998ahf,Sen:2015nph}. In particular, a democratic and manifestly $SL(2,\mb R)$ invariant type IIB action was constructed in the PST formulation in \cite{DallAgata:1998ahf}. An analogous democratic construction in the type IIA case was done in \cite{Bandos:2003et}.

We present here a new $O(10,10)$-adapted approach that treats all RR fields and their duals in a democratic manner \cite{DallAgata:1998ahf,Bergshoeff:2001pv} and is universal for both type II supergravities. We also derive an action for type IIB case with manifest $SL(2,\mb R)$ symmetry.

\section{Type II supergravity Action in the $O(10,10)$-adapted form}\label{sec:ingredients}
    We follow the exposition from \cite{Coimbra:2011nw} which uses the interpretation of the Ramond--Ramond fields in terms of $O(10,10)$-spinors \cite{Brace:1998xz}.
    
    We consider 10-dimensional spacetime with a metric of Lorentzian signature. In addition, we assume the presence of a closed 3-form flux $H$. In this setup, we can then define the following operations on the space of (inhomogeneous) differential forms.
    
    First, there is a natural pairing (Mukai pairing), valued in the top-degree differential forms, given by
    \begin{equation}
        (\alpha,\beta):=(-1)^{\left\lfloor \frac{\deg\alpha}{2}\right\rfloor}(\alpha\wedge\beta)^{\on{top}},
    \end{equation}
    where $\lfloor\;\cdot\;\rfloor$ denotes the integer part of a number, $\deg \alpha$ is the degree of the differential form $\alpha$ and $(\;\cdot\;)^{\on{top}}$ stands for the top form part.
    Then, we define the ``reflection'' operator $\star$ and the ``differential'' $D$ by
    \begin{align}
        \star\alpha=(-1)^{\left\lfloor \frac{\deg\alpha}{2}\right\rfloor+\deg\alpha}*\alpha,\qquad D\alpha=d\alpha+H\wedge \alpha,
    \end{align}
    where $*$ is the Hodge star for the Lorentzian metric.
    These operations satisfy the following easy-to-check properties:
    \begin{itemize}
        \item $\star^2=1$ ($\star$ serves as a natural replacement of $*$ in the context of Ramond--Ramond fields),
        \item the pairing $(\alpha, \star\beta)$ is symmetric in $\alpha$ and $\beta$ (this serves as a replacement for $\alpha\wedge*\beta$),
        \item $D^2=0$ ($D$ replaces the de Rham differential),
        \item $\int_M (\alpha,D\beta)=-\int_M(D\alpha,\beta)$ if $M$ has no boundary,
        \item $D(f\alpha)=fD\alpha+df\wedge\alpha$, for any function $f$.
    \end{itemize}
    
    If $B$ is a 2-form, we can define the operation $e^B$ by
    \begin{equation}
        e^B\alpha=\sum_n\tfrac1{n!} \underbrace{B\wedge\dots\wedge B}_{n}\wedge\,\alpha.
    \end{equation}
    Conjugating the operator $D=d+H$ by $e^B$ we then get
    \begin{equation}
        e^{B}(d+H)e^{-B}=d+(H-dB).
    \end{equation}
    Since $H$ is locally exact, we can always locally relate the operator $D$ to the ordinary $d$ in this way.
    
    In particular, suppose that $\alpha$ satisfies $D\alpha=0$. Locally, writing $H=dB$, we get $0=D\alpha=e^{-B}d(e^B\alpha)$, which implies that $e^B\alpha$ is closed and hence we can write $e^B\alpha=d\beta+c$, where $c$ is a constant \footnote{The presence of the constant reflects the fact that even though we are working locally, the zeroth (de Rham) cohomology group is still nontrivial.}. Thus
    \begin{equation}\label{eq:cohomology}
        \alpha=e^{-B}d\beta+e^{-B}c=D(e^{-B}\beta)+e^{-B}c.
    \end{equation}
    In particular, if $\alpha$ has no degree zero component and is $D$-closed, then it is also locally $D$-exact.

\subsection{Ramond--Ramond fields}
    In type II string theory, the (field strengths of the) Ramond--Ramond fields correspond to either purely even (in type IIA) or purely odd (IIB) inhomogeneous differential forms $F$ satisfying
    \begin{align}
    DF=0,\label{DF} \\ \star F=F \label{sF}.    
    \end{align}
    The self-duality condition \eqref{sF} ensures that only half of the degrees of freedom are present. This makes the contact with the original description of the RR sector, where in type IIA one has a 0-form (called the Romans mass), 2-form, and a 4-form, while in the IIB case one has a 1-form, 3-form, and an (anti-)self-dual 5-form. For simplicity, we will make the usual restriction and set the Romans mass to zero. (Otherwise we would run into some cohomological complications, stemming from the extra term in \eqref{eq:cohomology}.) We thus have
    \begin{align}
        F&=F_2+F_4+F_6+F_8+F_{10},\qquad \text{(IIA case)}\\
        F&=F_1+F_3+F_5+F_7+F_9.\phantom{\;\;}\qquad \text{(IIB case)}
    \end{align}
    
    Let us now describe the potentials for the Ramond--Ramond fields. First, choose a cover of $M$ by contractible open sets $\mc U_i$. Then the potentials are given by a collection of purely odd (in type IIA case) or purely even (IIB) differential forms $A_i$ on $\mc U_i$, which satisfy $DA_i=DA_j$ on overlaps $\mc U_i\cap\mc U_j$. The $DA_i$'s then glue together into a globally defined field strength $F$. When varying the potentials, we assume that on overlaps we have $\delta A_i=\delta A_j$, and so the variation corresponds to a globally defined form $\delta A$ --- correspondingly, we have $\delta F=D\delta A$. Similarly, the gauge transformations are given by $\delta A_i=Dc$, with $c$ a globally defined differential form.
    
    In particular, since $F=DA$, the equation \eqref{DF} is automatically satisfied. Our aim is now to derive the second equation \eqref{sF} from an action principle.

\subsection*{Remark}
    Let us briefly comment on the more conceptual viewpoint of the above constructions. First, differential forms can be regarded as spinors (or more precisely spinor half-densities) of $Spin(10,10)$, while purely even and purely odd forms correspond to chiral spinors. The pairing can be interpreted as the invariant spinor pairing. The operator $D$ corresponds to the \emph{generating Dirac operator} of \cite{AX,Severa:2017oew,Severa:2018pag}. The choice of Lorentzian metric breaks the group $Spin(10,10)$ down to $Spin(9,1)\times Spin(1,9)$. Self-dual and anti-self-dual differential forms (in the sense of \eqref{sF}) live in $(\mathbf{16},\overline{\mathbf{16}})$ and $(\overline{\mathbf{16}},\mathbf{16})$ of this subgroup. The reflection operator $\star$ coincides with the chirality operator for $Spin(9,1)\subset Spin(10,10)$ \cite{Rocen:2010bk}.

\subsection{The RR-sector action}
    
    We start from the following action for RR fields:
    \begin{eqnarray}
        S_R(A,R,a)=\tfrac12 \int_M \left[(F+aQ,\star(F+aQ))+2(F,aQ)\right],\label{SR}\qquad
    \end{eqnarray}
    where
    \begin{align}
        \qquad F=DA,\quad Q=DR,
    \end{align}
    $a$ is a scalar field, and $R$ is of the same type as $A$: a collection of even/odd forms for type IIB/IIA supergravity. $R$ does not have to be globally defined, similarly to $A$.
    The equations of motion are given by the vanishing of
    \begin{eqnarray}
        E_A:=&D[\star(F+aQ)+aQ],\\ E_R:=&D[a(\star(F+aQ)-F)],\\ E_a:=&((1-\star)(F+aQ),Q).
    \end{eqnarray}
    
    The action is invariant under the usual gauge transformations $\delta A=D\Lambda$ and $\delta R=D\Omega$. We also have the symmetry (analogous to PST formulation, see, e.g., \cite{Buratti:2019guq,Bandos:2020hgy})
    \begin{equation}
        \delta A=-ada\wedge\Sigma, \qquad \delta R=da\wedge \Sigma,\qquad \delta a=0.\label{symda}
    \end{equation}
    Finally, any shift of $a$ which does not violate the condition $(da)^2\neq 0$ (see below) can be compensated by a change in $A$ and $R$ such that the action remains invariant \cite{Bansal:2021bis,Avetisyan:2022zza}.

\subsection{Derivation of self-duality}

    The equations of motion in particular imply
    \begin{equation}
        0=E_R-aE_A=da\wedge(\star-1)(F+aQ).
    \end{equation}
    If $(da)^2\neq 0$ everywhere, this implies \footnote{For the 10d case discussed here, a quick argument for this claim can be made using the Remark from page \pageref{SR}. Namely, if $\beta$ is any anti-self-dual form and $e$ any 1-form, then $e\wedge\beta$ can be understood as the action of an element $e_++e_-\in Cl(9,1)\oplus Cl(1,9)\subset Cl(10,10)$ on an element $\beta$ of $(\overline{\mathbf{16}},\mathbf{16})$, where $\|e_+\|^2=-\|e_-\|^2=\tfrac12 g(e,e)$. Since $e_+ (\overline{\mathbf{16}},\mathbf{16})\subset (\mathbf{16},\mathbf{16})$ and $e_- (\overline{\mathbf{16}},\mathbf{16})\subset (\overline{\mathbf{16}},\overline{\mathbf{16}})$, we see that equation $e\wedge\beta=0$ implies that both $e_+\beta=0$ and $e_-\beta=0$. However, since $e_+^2=\|e_+\|^2$ in $Cl(9,1)$, writing $0=e_+(e_+\beta)=\|e_+\|^2\beta$ implies $\beta=0$ (assuming $e$ is not null).}
    \begin{equation}
        \star(F+aQ)=F+aQ.\label{F+aQ}
    \end{equation}
    Plugging this back in $E_A=0$ we get 
    \begin{equation}
        da\wedge DR=0\,.\label{daDR}
    \end{equation}
    This in turn implies that locally
    \begin{equation}
        R=D\Omega+da\wedge \Psi,
    \end{equation}
    where $\Omega$ and $\Psi$ are some (inhomogeneous) differential forms.
    Thus, $R$ is locally pure gauge and can be gauged away.
    Note, that the field $a$ also does not contain any physical degrees of freedom, as can be seen from its off-shell shift symmetry. \footnote{Notice that the shift transformation of the field $a$ affects also the field $R$. Still, $R$ can be gauged away on-shell for any chosen $a$ (the two gauge transformations do not commute, see \cite{Bansal:2021bis,Avetisyan:2022zza}).}
    Hence, if the spacetime is topologically trivial, all the degrees of freedom are encoded in $F=DA$, which satisfies the self-duality condition \eqref{sF}. If the topology is non-trivial, one still obtains the self-dual combination $F+aQ$, but there might be finitely many residual degrees of freedom associated to the topology.
    Note also, that the Lagrangian is invariant with respect to the symmetry \eqref{symda} only up to boundary terms \footnote{One can add a boundary term that would extend the gauge symmetry \eqref{symda} to the boundary, but it would spoil other gauge symmetries.}, and a global transformation to $aQ=0$ everywhere is not always allowed. Given that $F+aQ$ is invariant with respect to the transformation \eqref{symda} and closed due to \eqref{daDR}, it is more natural to think of the degrees of freedom being encoded in $F+aQ$ rather than $F$. A consequence of this will be discussed in the Conclusions.

\subsection{The full action}
The full pseudoaction for the bosonic sector of both type II supergravities can be given as (following the conventions of \cite{Coimbra:2011nw}):
\begin{align}
    \hat S&=S_{NS}+\hat S_R\,,\\
    S_{NS}&=\frac1{2\kappa^2}\int \sqrt{-g}\,
    e^{-2\varphi}\,\left(\mathcal R+4(d \varphi)^2
    -\frac1{12}H^2\right)\,,\\
    \hat S_R&=\pm \frac1{8\kappa^2}\int(F,\star F)  \,.\label{hSR}
\end{align}
Here $H$ is a field strength of the Kalb--Ramond 2-form field $B$, $\varphi$ is the dilaton, and $\mathcal R$ is the Ricci scalar for the metric $g$.
In the last expressions we have upper/lower sign for type IIA/IIB case.
In order to pass from the pseudoaction to the action we simply replace $\hat S_R$ \eqref{hSR} with $S_R$ \eqref{SR}. The full action is then given as:
\begin{align}
    S=\frac1{2\kappa^2} \int \left[\sqrt{-g}\,
    e^{-2\varphi}\,\left(\mathcal R+4(d \varphi)^2
    -\frac1{12}H^2\right)\right.
    \nonumber\\
    \pm\left.\frac18\left\{(F+aQ,\star(F+aQ))+2(F,aQ)\right\} 
    \right] \,.\label{O(10,10)adapted}
\end{align}
Here again, the upper/lower sign corresponds to type IIA/IIB supergravity.
In this formulation the $SL(2,\mb R)$ symmetry of type IIB supergravity is not explicit. We will proceed now to derive another form of the action, adapted to the $SL(2,\mb R)$ symmetry.

\section{Type IIB supergravity action in the $SL(2,\mb R)$-adapted form}
    
If in type IIB case we pass to the Einstein frame and rename (and rescale) the variables, we obtain the following pseudoaction \cite{Townsend:1996xj}. 
\begin{align}
    \hat S=\frac1{2 \kappa^2}\int \sqrt{-g}\left\{
    \mathcal R-2[(d\phi)^2+e^{2\phi}(d\ell)^2]
    -\frac1{3}e^{-\phi}H^2\right.
    \nonumber\\
    -\left.\frac1{3}e^{\phi}(H'-\ell H)^2-\frac1{60}M^2\right\}-\frac1{96\kappa^2}\int C\wedge H\wedge H'\,,
\end{align}
where the field content is now given by a metric $g$, two scalars $\phi$ and $\ell$, two 2-forms $B$ and $B'$ with curvatures $H$ and $H'$, respectively, and a 4-form $C$ with curvature $F=dC$. Finally, we set $M:=F+\frac12 (B\wedge H'-B'\wedge H)$.
The group $SL(2,\mb R)$ acts on the complex scalar $\tau:=\ell+ie^{-\phi}$ by fractional linear transformations; similarly $H'$ and $H$ form an $SL(2,\mb R)$-doublet. 

We can now introduce the new type IIB action
\begin{eqnarray}
    S=\frac1{2\kappa^2}\int \sqrt{-g}\left\{
    \mathcal R-2[(d\phi)^2+e^{2\phi}(d\ell)^2]
    -\frac1{3}e^{-\phi}H^2\right.
    \nonumber\\
    -\left.\frac1{3}e^{\phi}(H'-\ell H)^2 \right\}+S_{SD}\,.\qquad\quad
\end{eqnarray}
replacing only the sector of the pseudoaction containing the self-dual gauge field $C$ supplemented with interactions involving both 2-forms $B$ and $B'$ with ($Q=dR$):
\begin{eqnarray}
    S_{SD}=\frac1{2 \kappa^2}\int \left[(F+aQ)\wedge*(F+aQ)+2\,F\wedge aQ\qquad\right.\nonumber\\
    \left.-2\,(1+*)(F+aQ)\wedge X+X\wedge*X\right]\,,\qquad\label{eq:sl2sd}
\end{eqnarray}
where $X=\frac12(B\wedge H'-B'\wedge H)$ is the $SL(2,\mb R)$ symmetric Chern-Simons interaction contribution \cite{Bekaert:1999sq}. The equations of motion are given by the vanishing of
\begin{eqnarray}
        E_A:=&d[*(F+aQ)+aQ-(1-*)X],\\ E_R:=&d[a(*(F+aQ)-F-(1-*)X)],\\ E_a:=&Q\wedge(*-1)(F+aQ+X).
    \end{eqnarray}
In particular, we get
    \begin{equation}
        0=E_R-aE_A=da\wedge(*-1)(F+aQ+X).
    \end{equation}
Assuming $(da)^2\neq 0$ (almost) everywhere, we obtain 
\begin{equation}\label{eq:sl2selfduality}
    *(F+aQ+X)=F+aQ+X.
\end{equation}
Plugging this back in $E_A=0$ we get $da\wedge dR=0$ (here $R$ is only a four-form), which locally implies that
\begin{equation}
    R=d\omega+da\wedge\psi,
\end{equation}
for some 4-forms $\omega$ and $\psi$, and so we can gauge it away.

Let us check that the equations of motion for $B$ and $B'$ given by the new action coincide with the original ones. Varying \eqref{eq:sl2sd} w.r.t.\ $B$, we get a term
\begin{eqnarray}
    E_B=2H'\wedge[(1+*)(F+aQ)+*X]\qquad\nonumber\\+2d[B'\wedge ((1+*)(F+aQ)+*X)].
\end{eqnarray}
Using the self-duality \eqref{eq:sl2selfduality}, the vanishing of $Q$, and the identity $2H'\wedge X=B'\wedge dX$, this reduces to
\begin{equation}
    8H'\wedge (F+X).
\end{equation}
Similarly, the contribution to the equation of motion for $B'$ is
\begin{equation}
    -8H\wedge (F+X).
\end{equation}
This matches with the original variation, provided we choose the normalisation as in \eqref{eq:sl2sd}.

Note that the action \eqref{eq:sl2sd} is also gauge invariant under the familiar gauge transformations:
\begin{equation}
    \delta B=d\Lambda,\quad \delta B'=d\Lambda',\quad \delta C=\tfrac12(\Lambda'\wedge H-\Lambda\wedge H')\,.
\end{equation}
An important observation is that the self-duality equation can be amended by interactions with other fields via the mechanism put forward in \cite{Avetisyan:2022zza}, that is, adding to the Lagrangian a term $(1+*)(F+aQ)\wedge Y$, to modify the self-duality equation \eqref{eq:sl2selfduality} as:
\begin{eqnarray}
    (*-1)(F+aQ+X+Y)=0\,,
\end{eqnarray}
similarly to the addition of interactions with $B$ and $B'$ encoded in $X$.
This suggests an immediate way to incorporate the quadratic fermion interactions (the expression for the fermionic bilinear correction $Y$ to the self-duality equation is given, e.g., in \cite{Ciceri:2014wya}). 

\section{Conclusions}

We constructed a novel Lagrangian \eqref{O(10,10)adapted} for type II supergravities which is universal for both type IIA and type IIB. The fundamental fields of the RR sector are collections of odd/even forms for type IIA/IIB, described by a democratic Lagrangian using the formulation of \cite{Bansal:2021bis}. We also constructed a Lagrangian where only the self-dual field of the type IIB theory is treated as in \cite{Mkrtchyan:2019opf}, manifesting the $SL(2,\mb R)$ symmetry also present in the non-democratic pseudoaction. 
One interesting question would be to look for a formulation adapted to both $O(10,10)$ and $SL(2,\mb R)$ symmetries. This might require to reformulate the scalar and the two-form of the NS-NS sector in a democratic manner, introducing dual (eight- and six-form) potentials for them.

As discussed, e.g., in \cite{DallAgata:1998ahf,Bandos:1997ui}, the democratic formulation is useful also for coupling branes to the supergravity.

One can perform dimensional reduction of the actions presented here to arrive at novel democratic Lagrangians of maximal supergravities in different lower dimensions $d=11-n\; (n>1)$, with $E_{n(n)}$ duality symmetries (see, e.g., \cite{Hull:2007zu,PiresPacheco:2008qik}, \cite{Bossard:2021jix} and references therein).

Another application of the actions presented here would be to use the extra symmetries involved in them compared to the pseudoactions (together with supersymmetry) to constrain higher-order $\alpha'$-corrections to the string effective action (see \cite{Coimbra:2014qaa,Liu:2022bfg} and references therein). The unique quartic invariant of the self-dual four-form field in ten dimensions was already identified in \cite{Avetisyan:2022zza} (see also \cite{Buratti:2019guq} where the quartic vertex in the perturbative expansion was found in PST formulation), allowing to define a large class of full non-linear theories of self-interacting chiral four-form (see also \cite{Evnin:2022kqn}). These findings can have a direct application to the type IIB case. Higher-order interactions are less constrained for a single chiral form, but can be much more constrained given the $O(10,10)-$adapted structure works at higher orders.

An interesting aspect of type IIB supergravity was discussed in \cite{Kurlyand:2022vzv}. A resolution of the puzzle related to the on-shell value of the action on product manifolds was suggested, which is related to a boundary term, specific to the background. Another potential resolution can be derived from the Lagrangian \eqref{eq:sl2sd}. There, a solution \footnote{We thank Arkady Tseytlin for discussion on this point and sharing with us their notes with Stephan Kurlyand.} of the equations of motion with a background spacetime $M^5\times X^5$ can be given by $F$ being proportional to the volume form of $M^5$ and $aQ$ being proportional to the volume form of $X^5$, with the same coefficient, so that $F+aQ$ is self-dual, solving the equation \eqref{F+aQ}. Then, the on-shell value of the action is given by the second term $F\wedge aQ$ in the Lagrangian \eqref{eq:sl2sd} and is proportional to the volume of the space-time. Note, that $aQ$ is closed on-shell and therefore $F\wedge aQ$ becomes a boundary term on-shell, satisfying the same conditions as the topological term added in \cite{Kurlyand:2022vzv}, except that now this term is already encoded in the 10d covariant action and is not background-specific.

\vspace{.2cm}
\noindent The authors are grateful to Zhirayr Avetisyan, Franz Ciceri, Oleg Evnin, Axel Kleinschmidt, Arkady Tseytlin and Daniel Waldram for helpful discussions on the subject of this work. The authors are also indebted to Augusto Sagnotti for the encouragement to look into this problem, and to Ivano Basile, Amihay Hanany, Salvatore Raucci and Fiona Seibold for related discussions. K. M. is supported by the
European Union’s Horizon 2020 Research and Innovation
Programme under the Marie Skłodowska-Curie Grant
No.\ 844265 and in part by the STFC Consolidated Grant ST/T000791/1.  F. V. was supported by the Postdoc Mobility grant P500PT\underline{\phantom{k}}203123 of the Swiss National Science Foundation.

\end{document}